\documentclass[aps,prb,reprint,superscriptaddress,longbibliography,floatfix]{revtex4-2}
\usepackage[colorlinks=true,linkcolor=blue,citecolor=blue,urlcolor=blue]{hyperref}
\usepackage{microtype} 
\usepackage{graphicx} 
\usepackage{amsfonts, amsthm, amsmath, amssymb, physics, upgreek, stackrel,bm} 
\usepackage{array, enumitem} 
\usepackage{verbatim, listings} 
\usepackage{dsfont, xr, color, lipsum} 
\usepackage{placeins} 

\usepackage[x11names,dvipsnames]{xcolor}
\usepackage{JanShortcuts}

\definecolor{StrangeGreen}{HTML}{58B359}
\usepackage{orcidlink}
\newcommand{\orcidLouw}{\orcidlink{0000-0002-5111-840X}}

\begin{document}

\title{What does ``instant thermalization'' in large-$q$ SYK models mean?}
\author{Alexander Osterkorn}
\affiliation{Jo\v{z}ef Stefan Institute, SI-1000, Ljubljana, Slovenia}
\author{Jan C. Louw \orcidLouw}
\email{jancillie.louw@gwdg.de}
\affiliation{Arbeitsgruppe Computing (AG C), Gesellschaft f\"ur wissenschaftliche Datenverarbeitung mbH G\"ottingen (GWDG), Burckhardtweg 4, D-37077 G\"ottingen, Germany}


\begin{abstract}
Motivated by the Planckian thermalization rate $\Gamma \sim T$ observed in the two-body interacting SYK model, we study thermalization in the $q/2$-body case.
Previous studies of such systems have established the notion of instantaneous thermalization to leading order in $1/q$.
It was conjectured that the thermalization may still be Planckian but with a divergent rate $\Gamma \sim T q$, explaining the ``instantaneous'' part.
For an analytically and numerically tractable system, we calculate the effective temperatures after a quench at time $t = 0$ and indeed find a Planckian rate, albeit with the unexpected decaying behavior $\Gamma \sim T q^{-1} $. This is contrasted with the behavior of the causal Green's function $\Gg(t_1, t_2)$, which instantly acquires a thermal form in the two-time plane block $t_1, t_2 > 0$.

The resulting picture is that instant thermalization is a meaningful concept only for this block, while the off-diagonal blocks $t_1 \cdot t_2 < 0$ are inherently non-thermal at any $q$. Since the effective temperature is obtained from correlations spanning all blocks, it inherits a finite rate set by the off-diagonal. We illustrate this directly by studying the non-thermal correlations' time dependence.
\end{abstract}

\maketitle

\section{Introduction}

Understanding thermalization in strongly interacting quantum systems remains a central problem in nonequilibrium many-body physics. The Sachdev-Ye-Kitaev (SYK) model \cite{SachdevYe1992,Kitaev2015,Chowdhury2022Sep} provides a particularly useful setting, as it combines solvability with features characteristic of chaotic quantum matter. As such, the problem has been studied extensively \cite{Eberlein2017Jun,Louw2022Feb,Almheiri2024Aug,Sohal2022Jan,Halataei2025Jul,Jaramillo2025May,Perugu2025Nov,Bhattacharya2019Jul,Bandyopadhyay2023May}. Often one focuses on $q/2$-body SYK type interactions, since the model is exactly solvable at leading order in $1/q$. In equilibrium, both finite-$q$ and large-$q$ versions of the model exhibit similar thermodynamic \cite{Azeyanagi2018Feb,Ferrari2019Jul,Louw2023Feb} and transport properties \cite{Zanoci2022Apr,Jha2025Jan}, including saturating the Planckian bounds on chaos \cite{Maldacena2016Aug}. 

Out of equilibrium, however, the connection between finite-$q$ and large-$q$ dynamics is less understood. Especially for quenches the two cases seem to fall into completely different classes of thermalization. It is known that the standard $q=4$ SYK model thermalizes at a Planckian (universal and linear in temperature) rate $\Gamma \sim k_{B} T_f/\hbar$ \cite{Eberlein2017Jun} for low final temperatures $T_f$. This rate is also associated with black hole equilibration \cite{Lei2022Apr,Vishveshwara1970Aug,Carullo2021Apr,Louw2023Oct}. Specifically, this rate was seen in the relaxation of the effective temperature defined via a non-equilibrium (NEQ) fluctuation dissipation relation.
Further, taking $q \to \infty$ one can analytically show that to leading order in $1/q$, the post-quench two-point correlation functions immediately satisfy a KMS relation. In this way the system is said to obey instant thermalization~\cite{Eberlein2017Jun,Louw2022Feb}. To our knowledge, the SYK model is the only known model to exhibit this behavior.

It was conjectured in \cite{Eberlein2017Jun} that these two regimes may be connected through a diverging rate of the form $\Gamma \propto q T_f$. Such a scaling appears natural if one interprets $q$ as the effective connectivity of the system, following the perspective of~\cite{Goldstein2015Apr}. A direct distance measure to equilibrium provides a lower bound on thermalization rate \cite{Nickelsen2019May} which indeed grows with $q$. The full bound unfortunately tends to zero in the thermodynamic limit; thus, it is not applicable to our SYK system where this limit is taken. The typical random Haar evolution rate is again found to be Planckian \cite{Goldstein2015Apr}. This Planckian rate is the so-called bound on chaos and was initially shown to be the maximal decay rate of out-of-time-order correlation (OTOC) functions \cite{Maldacena2016Aug}. Later it was also demonstrated to be a bound on the decay rate for thermal two-point correlation functions $\Gg_{\text{eq}}(\tau)$ \cite{Pappalardi2022Apr}. It is \emph{not} however proven to be an upper bound on the thermalization rate.

Against the backdrop of these results, a more detailed analysis is needed that contrasts different concepts of thermalization and specifically the notion of instant thermalization. In this paper, we therefore focus on two established indicators:
(i) the equilibration of Green's functions,  
(ii) the relaxation of effective temperatures. 
Concretely, we consider quenches between multiple non-commuting $q/2$-body SYK Hamiltonians. Our setup is tractable both numerically and analytically allowing for an in-depth analysis. The analytic results are demonstrated in a companion paper~\cite{CompanionLouw2026}. 
Specifically, we can even provide analytical solutions for the inherently non-thermal off-diagonal time-blocks ($t_{1} t_{2} < 0$) where operators evolve under different Hamiltonians across the quench occurring at $t=0$. This distinguishes it from the quenches in \cite{Eberlein2017Jun,Louw2022Feb} for which only the thermal blocks are known in closed form, and have not been directly connected to another. The present approach provides access to the entire time plane, as sketched in Fig.~\ref{fig:blocks}, enabling a more complete picture of the relaxation process, and hence a structured comparison between different thermalization scenarios.

We proceed as follows: in Sec.~\ref{secGdiff} we find that up to order $1/q^{2}$ the finite-$q$ Green's functions reproduce the large-$q$ solution over the whole two-time plane, which would be consistent with a diverging rate $\Gamma$. In Sec.~\ref{secTeff} we find the opposite $\Gamma \sim T_f q^{-1}$. Sec.~\ref{secblocks} shows both to hold, since the two measure distance to different reference states: the former compares finite $q$ against large $q$, the latter against thermality, which the large-$q$ solution obeys only in the diagonal time-blocks. Extracting a rate from the off-diagonal blocks directly yields a third answer, $2\pi/\tau$, set by the relative time $\tau \equiv t_1-t_2$ probed rather than by $T_f$ or $q$. As such each notion of thermalization carries its own rate, and ``instant thermalization'' is to be read as a statement about the diagonal time-blocks alone.

\section{Model and Setup} \label{secSetup}
We consider a single SYK dot of $N$ spinless complex fermions subject to two mutually non-commuting $q/2$-body interactions with time-dependent strengths,
\begin{equation}
    H(t) = \sum_{i=1}^{2}\, \hspace{-1mm} \sum\limits_{\substack{ \{\bm{\mu}\}_1^{q/2} \\ \{\bm{\nu}\}_1^{q/2} }} \hspace{-2mm}Z^{(i)}(t)^{\bm{\mu}}_{\bm{\nu}} \hat{c}^{\dag}_{\mu_1} \cdots \hat{c}^{\dag}_{ \mu_{q/2}} \hat{c}_{\nu_{q/2}}^{\vphantom{\dag}} \cdots \hat{c}_{\nu_1}^{\vphantom{\dag}},
    \label{hi}
\end{equation}
where $\{\bm{\nu}\}_{1}^{q/2} \equiv 1\le \nu_1<\cdots< \nu_{q/2}\le\Nn$, and $\hat{c}_{\alpha}^{\dag}, \hat{c}_{\alpha}^{\vphantom{\dag}}$ create and annihilate a fermion of flavor $\alpha$. The couplings $Z^{(i)}(t)^{\bm{\mu}}_{\bm{\nu}}$ are independent Gaussian random variables with zero mean and variance  $\overline{|Z^{(i)}|^2} = (N/2)^{1-q}[(q/2)!\Jj^{(i)}/q]^2$, drawn independently for each $i$. This setup can also be used to describe a uniform chain with a $q/2$-body transport term \cite{Jha2023Jun,Jha2025Jan}. Our protocol can be summarized by grouping the effective coupling strengths into the vector 
\begin{equation}
    \vec{\Jj}(t) = (\Jj^{(1)},\Jj^{(2)}) = ( \Jj,\Jj + \Jj\gamma\Theta(t)),
\end{equation}
i.e., we only change the second Hamiltonian over the quench at time $t=0$. Without loss of generality, we will set $\Jj=2$ throughout this work which just sets an overall energy scale. The parameter $\gamma$ however tunes the strength of the quench. Intuitively larger $\gamma$ should result in a higher final temperature. The results are qualitatively independent of $\gamma$; so we simply set $\gamma = 1/2$.

Remarkably in the thermodynamic limit ($N\to\infty$) the SYK model is solvable in the sense that one may obtain the exact self-energy
\begin{align}
    \Sigma(t_1,t_2) =& -2\vec{\Jj}(t_1) \cdot \vec{\Jj}(t_2)/q \notag\\ & \cdot[4 \Gg(t_1,t_2)\Gg(t_2,t_1)]^{q/2-1}\Gg(t_1,t_2). \label{self}
\end{align}
The fact that the two Hamiltonians are non-commuting is reflected in the lack of factorization in $\vec{\Jj}(t_1) \cdot \vec{\Jj}(t_2) \neq \sum_{i}\Jj^{(i)}(t_1)\sum_{j}\Jj^{(j)}(t_2)$. In the case of a single SYK Hamiltonian, only trivial ``quenches'' (the factorizing type) can be considered, since the time-dependent coupling merely amounts to a change in energy scale rather than any change in the system. 

The self-energy \eqref{self} corresponds to a multitude of SYK Hamiltonians under specific conditions. The important feature shared among all interpretations is that it consists of two or more non-commuting $q/2$-body SYK interacting terms. As such, a quench of one coupling actually yields non-trivial physics. This contrasts with the setup where a commuting term is quenched, for instance, the mass term or the case with a single SYK term with a time-dependent coupling constant. The latter is merely related to a change in units; hence it leaves equilibrium physics undisturbed.

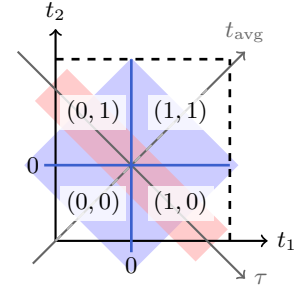
\begin{figure}
    \centering
    \begin{tikzpicture}[scale=1]
\def\T{1}
\def\R{1}

\fill[blue!40, fill opacity=0.5, rotate=45] (-1*\T, -1*\T) rectangle (1*\T, 1*\T);
\fill[red!40, fill opacity=0.5, rotate=-45] (-1.5*\T, -0.3*\T) rectangle (1.5*\T, 0.3*\T);

\draw[->, thick] (-1*\T,-1*\T) -- ({(\R+0.8)*\T},-1*\T) node[right] {$t_1$};
\draw[->, thick] (-1*\T,-1*\T) -- (-1*\T,1.8*\T) node[above] {$t_2$};

\draw[->, thick, black!60] (-1.3*\T, -1.3*\T) -- (1.5*\T, 1.5*\T) node[above] {$t_{\text{avg}}$};
\draw[->, thick, black!60] (-1.5*\T, 1.5*\T) -- (1.5*\T, -1.5*\T) node[right] {$\tau$};

\draw[black!30, thick, dashed] (-1*\T,-1*\T) -- (1*\T,1*\T);

\draw[RoyalBlue3, line width=1pt] (0*\T,-1.15*\T) -- (0*\T,1.4*\T);
\draw[RoyalBlue3, line width=1pt] (-1.15*\T,0*\T) -- (1.3*\T,0*\T);

\draw[black, line width=1pt, dashed] (1.3*\T,-1*\T) -- (1.3*\T,1.4*\T);
\draw[black, line width=1pt, dashed] (-1*\T,1.4*\T) -- (1.3*\T,1.4*\T);

\node[fill=white, fill opacity=0.8, text opacity=1, inner sep=2pt] at (-0.5*\T,-0.5*\T) {$(0,0)$};
\node[fill=white, fill opacity=0.8, text opacity=1, inner sep=2pt] at (0.65*\T,-0.5*\T) {$(1,0)$};
\node[fill=white, fill opacity=0.8, text opacity=1, inner sep=2pt] at (-0.5*\T,0.7*\T) {$(0,1)$};
\node[fill=white, fill opacity=0.8, text opacity=1, inner sep=2pt] at (0.65*\T,0.7*\T) {$(1,1)$};

\node[below] at (0,-1.1*\T) {$0$};
\node[left] at (-1.1*\T,0) {$0$};

\end{tikzpicture}
    \caption{The four blocks of the two-time plane, labelled by the pair of Hamiltonians $(s_1,s_2)$ under which the two operators are evolved, with $0$ and $1$ denoting pre- and post-quench. The two diagonal time-blocks $(0,0)$ and $(1,1)$ are thermal, at $T_i$ and at $T_f$ respectively, the latter to leading order in $1/q$ and, at finite $q$, up to corrections of order $1/q^2$ (Fig.~\ref{fig:dq}), whereas the off-diagonal time-blocks $(1,0)$ and $(0,1)$ are inherently non-thermal at any $q$. Also shown are the Wigner coordinates ($\tau, t_{\text{avg}}$) \eqref{wigner_co}.  We include two tilted rectangles used for performing the Fourier transform with respect to $\tau = t_1-t_2$.}
    \label{fig:blocks}
\end{figure}

We study the quench dynamics of the system by focusing on the  NEQ (time ordered or causal) Green's functions $\Gg(t_1,t_2) \equiv -\imath \expval{\Tt \hat{c}^{\vphantom{\dag}}(t_1) \hat{c}^\dag(t_2)}$ averaged over all $N$ fermionic flavors. This can be explicitly written in terms of the lesser and greater components
\begin{align}
   \Gg(t_1,t_2) &= \Theta(t_1-t_2)\Gg^{>}(t_1,t_2) + \Theta(t_2-t_1)\Gg^{<}(t_1,t_2) \notag\\
   &= \Gg^\text{ret}(t_1, t_2) + \Gg^<(t_1, t_2). \label{time_orderedG}
\end{align}
For $t_1 > t_2$, this is equivalent to $\Gg^>(t_1, t_2)$ by means of the identity $\Gg^\text{ret}(t_1, t_2) = \Theta(t_1-t_2) \big( \Gg^{>}(t_1,t_2) - \Gg^{<}(t_1,t_2) \big)$.
With the SYK set-up described above, one is able to solve for the thermal pre-quench ($t_1,t_2<0$) and post-quench time-block ($t_1,t_2>0$) \cite{Eberlein2017Jun,Louw2022Feb}. What one also needs is access to the blocks ($t_1<0,t_2>0$) and ($t_1>0,t_2<0$), which are inherently NEQ since the operators there are evolved under two different Hamiltonians, namely the pre-quench and post-quench ones. We denote the pre- and post-quench quantities with subscripts $0$ (or $i$) and $1$ (or $f$) respectively, as indicated in Fig.~\ref{fig:blocks}. We make use of a novel solution to these off-diagonal time-blocks, derived in the companion paper~\cite{CompanionLouw2026}, partially restated in App.~\ref{fullSol}. 

\section{Thermalization Measures} \label{secMeasures}

Green's functions in equilibrium satisfy a KMS relation---a periodicity $\beta$ in imaginary time---meaning the frequency sets the temperature $1/\beta = T$.
Consequently, observables typically carry characteristic dependencies on temperature, even beyond equilibrium.
However, in our specific zero-dimensional system, the only meaningful observables we have access to are conserved quantities: the energy and the number density. Therefore, we will turn our attention to the Green's function itself.
Full numerical solutions for Green's functions with the self-energy~\eqref{self} and varying values of $q$ have been obtained using the NESSi library~\cite{Schueler2020}.
For our quench setup, we consider two natural measures to monitor the thermalization process: The first is to have some measure of how far away the Green's functions are from a thermal state~\cite{Goldstein2015Apr}, and the second is the change of an effective temperature~\cite{Eberlein2017Jun}.

\subsection{Real-Time Green's function} \label{secGdiff}
 In the following, we study the distance of the nonequilibrium Green's functions away from the analytical large-$q$ prediction $\Gg^{(\text{ana})}$ directly in the two-time plane, meaning over all four time-blocks. Note that $\Gg^{(\text{ana})}$ is instantaneously thermal only in its diagonal time-blocks; in the off-diagonal ones it is the inherently non-thermal solution of App.~\ref{fullSol}.
Our solution is derived under a $1/q$ expansion which is re-summed such that it overlaps with the finite-$q$ low temperature conformal solution as well as the standard large-$q$ solution discussed more in App.~\ref{appResum}.
It is this low-temperature regime that is our main regime of interest. We focus on the Green's functions as a function of time difference and time average 
\begin{equation}
\tau = t_1 - t_2 \quad t_\text{avg} = (t_1+t_2)/2 , \label{wigner_co}
\end{equation}
corresponding to Wigner coordinates.
While the conformal solution is valid at all $q$, it is only strictly correct at late time differences $\tau$.
As such, we expect deviations from the numerical result at intermediate $\tau$ even if we formally take the zero temperature limit.

\begin{figure}
 \centering
 \includegraphics[width=0.9\linewidth]{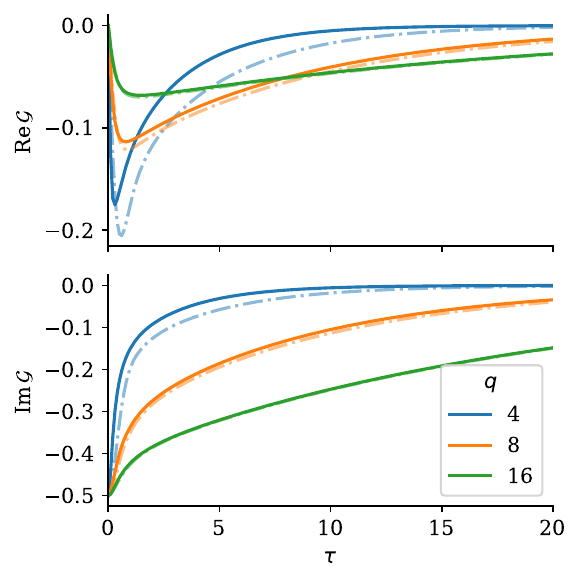}
 \caption{The real and imaginary parts of the NEQ time ordered Green's function \eqref{time_orderedG} as a function of relative time $\tau$ directly after the quench (sharp ramp $t_{\text{avg}} = 10^{-3}$). The data is for $T_i = 0.1$ and various connectivities $q$. The dashed-dotted line corresponds to the numerical results, while the analytical prediction $\Gg^{(\text{ana})}$ is given by the solid line of the same color.\label{fig:placeholder}. Since $\tau>0$, the data used here is that of $G^>(t_1,t_2)$.}
\end{figure}

Focusing on the Green's function at times close to the quench, we plot real and imaginary parts of its full numerical solution $\Gg(t_1,t_2)$ and of the instantaneous thermal prediction $\Gg^{(\text{ana})}(t_1,t_2)$ from the large-$q$ result in Fig.~\ref{fig:placeholder}.
Importantly, we observe that the numerical solution is close to the large-$q$ prediction even directly after the quench.
Note that the Green's functions overlap almost perfectly at $q = 16$.
In other words, the difference between the full numerical solution and the analytical prediction tends to zero as we increase $q$.
This may be expected given that the analytic solution is derived under a $1/q$ expansion.

\begin{figure}[h!]
    \centering
    \includegraphics[width=0.8\linewidth]{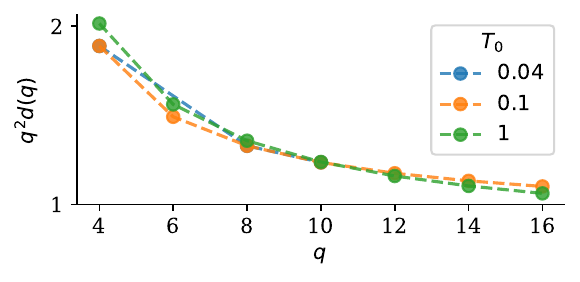}
    \caption{Maximum difference between Green's functions \eqref{eq:def_dq} multiplied by $q^2$ as a function of the connectivity $q$ for three different initial temperatures $T_0$. 
    }
    \label{fig:dq}
\end{figure}

For a more quantitative assessment, following Ref.~\cite{Goldstein2015Apr}, we choose a distance measure between the full numerical NEQ Green's function and the analytical result 
\begin{equation}
d(q) = \max\limits_{\substack{t_1,t_2}}
    |\Gg(t_1,t_2)-\Gg^{(\text{ana})}(t_1,t_2)|,
    \label{eq:def_dq}
\end{equation}
as a function of $q$.
In Fig.~\ref{fig:dq}, we directly compare the two solutions by considering the difference measure \eqref{eq:def_dq} multiplied by $q^2$ for various temperatures $T$.
The temperatures are chosen to be small enough such that we are in the Planckian range, $T \lesssim \Jj/2$, i.e., where the Lyapunov exponent is close to maximal.

The data in Fig.~\ref{fig:dq} strongly suggests that $q^2 d(q)$ tends to a constant as $q \rightarrow\infty$, which implies that deviations between the numerical solution of the Green's function and the large-$q$ theory are of the order $1/q^2$.
This is the typical behavior for all temperatures in the Planckian range, e.g., those temperatures considered in Fig.~\ref{fig:dq}.
As a consequence, even the exact finite-$q$ Green's functions are reproduced by the large-$q$ solution at order $1/q$, and are therefore instantaneously thermal at that order wherever the large-$q$ solution is, namely in the diagonal time-blocks.  

In summary, the finite-$q$ Green's functions overlap strongly with the large-$q$ solution not only at large $q$, but significantly so already at $q=4$. Naively this can lead one to conclude that $\Gamma$ lies beyond any finite scale, so that the conjecture $\Gamma \propto q T_{f}$ of Ref.~\cite{Eberlein2017Jun} would be an understatement rather than an overstatement. We next contrast the above with a second indicator.

\subsection{Effective Temperature} \label{secTeff}
We now contrast the real-time picture in Sec.~\ref{secGdiff} with a spectral indicator of thermalization:
Effective temperatures $T_\text{eff}$ are defined on the basis of the fluctuation-dissipation theorem for correlations in equilibrium.
The equilibrium spectral function $A(\omega) = G^<(\omega) \big( 1 + \text{e}^{-\beta\omega} \big)$ and the Keldysh Green's function in Keldysh space, $G^{\text{K}}(\omega) = G^>(\omega) + G^<(\omega)$ combine to
\begin{align}\begin{split}
 i G^{\text{K}}(\omega) / A(\omega) = \tanh(\beta\omega/2) \simeq \beta\omega/2 ,
\end{split}\end{align}
where the last statement holds for $\beta\omega/2 \ll 1$.
Following the work by Eberlein et al.~\cite{Eberlein2017Jun}, we also calculate this ratio for Green's function out of equilibrium, formulated with Wigner time coordinates~\eqref{wigner_co}.
We evaluate the above mentioned ratio of the resulting nonequilibrium Green's functions $G(t_\text{avg}, \omega)$ and fit an effective temperature from its low-frequency behavior.
A detailed discussion of our data processing and sources of error can be found in Appendix~\ref{app:numerics}.
Fig.~\ref{fig:efftemps_gammas}(a) and (b) exemplarily show results of this procedure both for the full numerical solution and for the analytical large-$q$ theory.
Since the large-$q$ theory reaches thermal equilibrium instantaneously in the diagonal block, one might assume the curve would need to be a step function. However, we find that neither analytic nor numerical solutions behave like a step function.
A reason for this is that the off-diagonal time blocks $t_1 \cdot t_2 < 0$, which we have not discussed in detail before, carry memory of the pre-quench dynamics and contribute to the low-frequency behavior of the Green's function.
Note that this is unavoidable by construction: $T_\text{eff}(t_\text{avg})$ is read off from the transform along the whole line $t_{1,2} = t_\text{avg} \pm \tau/2$, which leaves the post-quench time-block as soon as $\vert \tau \vert > 2 t_\text{avg}$, and the low-frequency fit weights precisely the largest $\tau$.

We find that the overall shapes of the effective temperature curves obtained with the two approaches (numerical and analytical) coincide well and indicate a finite thermalization rate.
We further note that the large-$q$ theory, especially for $q = 4$, underestimates the final temperature established after the quench. We can, however, see that this underestimation appears to decrease as we increase $q$, as expected from Fig.~\ref{fig:dq}.

\begin{figure}
\includegraphics[width=0.42\textwidth]{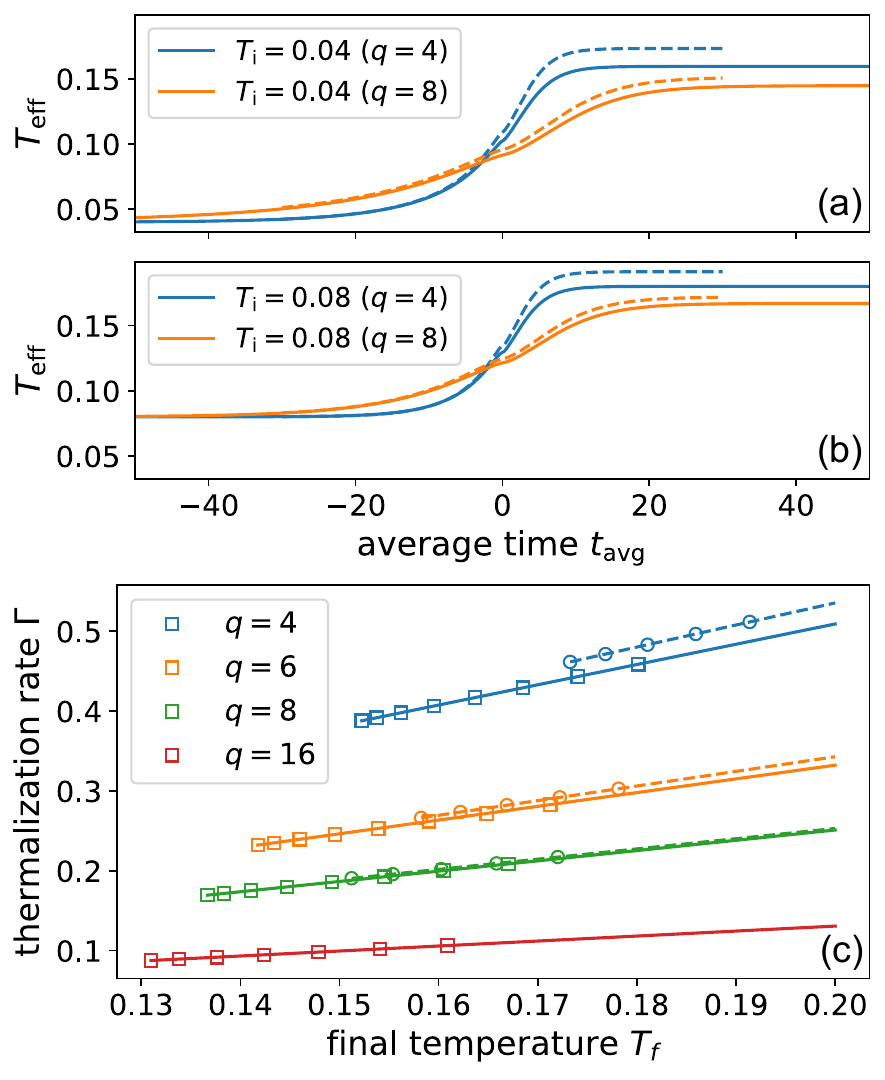}
\caption{(a,b) Effective temperatures from the numerical solution (dashed lines) and from the large-$q$ theory (solid lines) for two different initial temperatures $T_i = 0.04, 0.08$ and for $q = 4, 8$.
(c) Thermalization rates $\Gamma$ extracted from the effective temperatures of the numerical solution (circles, dashed) and of the large-$q$ theory (squares, solid).
The lowest considered initial temperatures are $T_i = 0.04$ for the numerical solution and $T_i = 0.01$ for the large-$q$ solution.\label{fig:efftemps_gammas}}
\end{figure}

\begin{figure}
\includegraphics[width=0.4\textwidth]{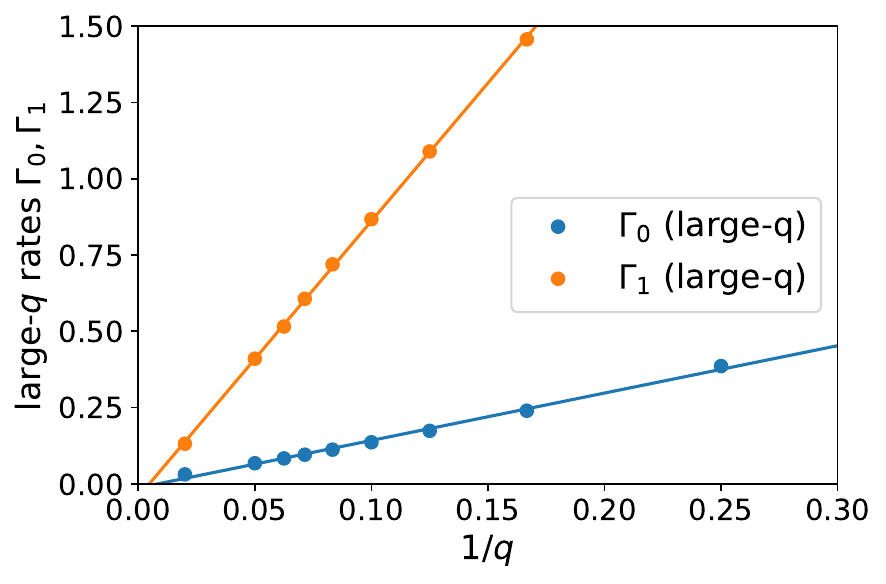}
\caption{Dependence of the large-$q$ thermalization rates $\Gamma_0, \Gamma_1$ (as defined in \eqref{eq:def_therm_rates}) on $q$. The plotted thermalization rate $\Gamma_1$ is evaluated at the smallest considered temperature $T_i = 0.01$.
The solid lines are a linear fit to the data and provide a guide to the eye showing that the data scales roughly linearly with $1/q$.
\label{fig:gammas_vs_qs}}
\end{figure}

Thermalization rates at late average times after the quench are extracted using an ansatz
\begin{align}\begin{split}
 T_\text{eff}(t_\text{avg}) = T_f + \alpha \text{e}^{-\Gamma t_\text{avg}} .
\end{split}\end{align}
In Fig.~\ref{fig:efftemps_gammas}(c) we present the extracted thermalization rates for the full numerical ($q = 4, 6, 8$) and for the large-$q$ solution, for which we can more easily cover larger values of $q$.
In both cases, the rates depend linearly on the final temperature $T_f$ in agreement with the Planckian rates observed in~\cite{Eberlein2017Jun} for $q=4$.
The low-temperature end point of the curve is set by the lowest possible final temperature $T_f^\text{min}$, which corresponds to the temperature obtained after quenching from an initial state with temperature zero.
$T_f^\text{min}$ moves to lower values as $q$ increases.
The data also clearly shows that the linear-in-$T_f$ parts of the thermalization rates obtained from the full and large-$q$ solutions become increasingly similar at larger values of $q$.
Since the smallest initial temperature $T_i$ used for our full numerical solution is $T_i = 0.04$, the low-temperature end points of the data cannot be directly compared.
However, comparing data points with the same value of $T_i$, we can also find that they approach each other as $q$ increases.
Looking at the values of the rates $\Gamma$, they clearly seem to decay with $q$, which is in stark contrast to the hypothesized diverging-in-$q$ rates discussed in \cite{Eberlein2017Jun}.
For a more quantitative analysis, we split the thermalization rate $\Gamma$ into a constant part $\Gamma_0$, depending only on $T_f^\text{min}$, and a part $\Gamma_1$ describing the linear dependence on the final temperature:
\begin{align}\begin{split}
 \Gamma = \Gamma_0[T_f^\text{min}] + \Gamma_1 \cdot ( T_f - T_f^\text{min} ) .
 \label{eq:def_therm_rates}
\end{split}\end{align}
Fig.~\ref{fig:gammas_vs_qs} shows the extracted rates obtained within large-$q$ theory as a function of $1/q$ down to $q = 50$.
$\Gamma_1$ is evaluated for quenches with an initial temperature $T_i = 0.01$.
The solid straight lines are fitted from the data and provide a guide to the eye.
However, since this linear extrapolation of the data would predict slightly negative, unphysical limit values of $\Gamma_i$ as $q \rightarrow \infty$,
we cannot make a reliable statement about these precise values within the accuracy of our approach.
Nevertheless, this discussion based on effective temperatures very clearly does not support instant thermalization in the sense of increasing the thermalization rate as $q \rightarrow \infty$.

\subsection{Relaxation in the off-diagonal time-blocks} \label{secblocks}

\begin{figure}
    \centering
    \includegraphics[width=0.85\linewidth]{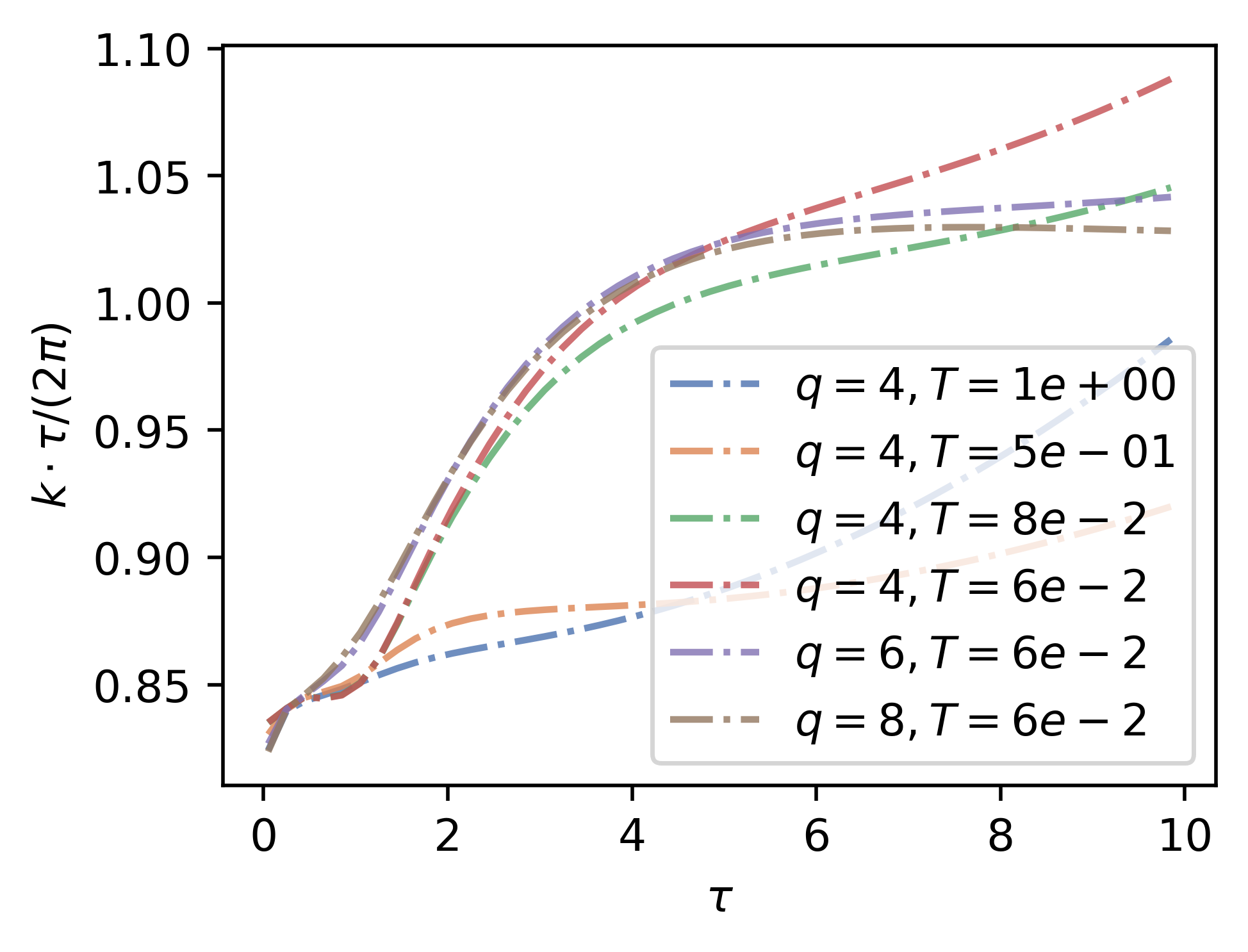}
    \caption{The behavior of the rate $k$ \eqref{k_rate} for different sets of numerical data. Note the insensitivity to $T,q$.}
    \label{kfit}
\end{figure}

The two subsections above measure different notions of thermality. The measure \eqref{eq:def_dq} of Sec.~\ref{secGdiff} shows that the finite-$q$ solution is faithfully captured by the large-$q$ one over the whole two-time plane. The thermal distance however appears only in $\Gg^{(\text{ana})}$ restricted to the diagonal time-blocks of Fig.~\ref{fig:blocks}. The effective temperature of Sec.~\ref{secTeff}, by contrast, measures the distance from thermality via a non-equilibrium fluctuation dissipation relation, which necessarily includes the off-diagonal time-blocks.
In order to shed more light on the role of the off-diagonal time blocks, we discuss a third focused measure: the difference of the full numerical solution $\Gg(t_1,t_1-\tau)$ and its late-time asymptotic (thermal) solutions $\Gg_{\text{eq}}(\tau)$.
\begin{figure}
    \centering
    \includegraphics[width=0.9\linewidth]{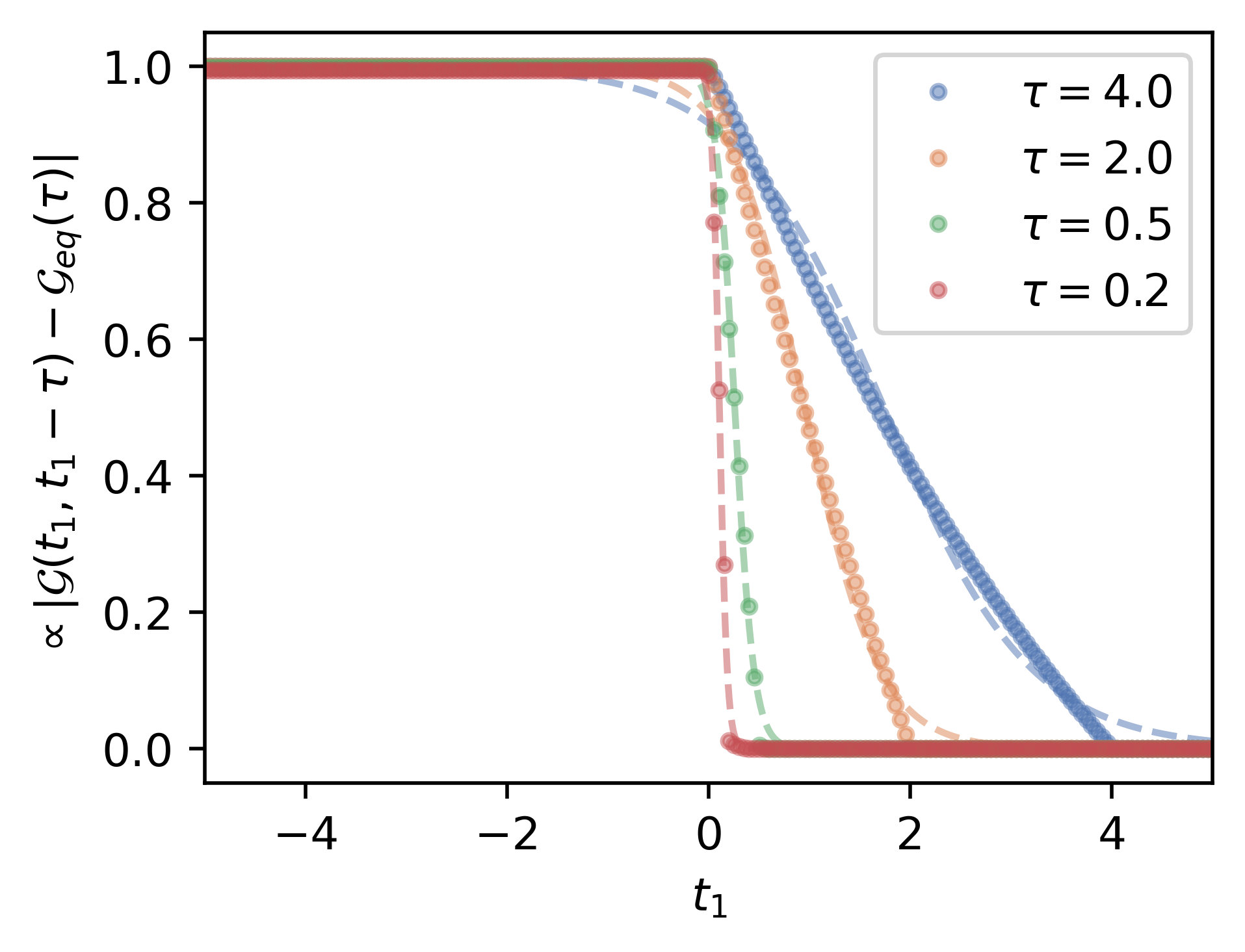}
    \caption{Plot of the time ordered Green's functions' \eqref{time_orderedG} difference \eqref{sigmoid}, scaled (note the $\propto$ symbol in the vertical label) such that the maximum falls onto $1$. The typical behavior is that of a sigmoid, shown here given here for $T = 0.5$ and $q=4$ and various relative times $\tau$. The numerical data is given in the dots, while the corresponding fit is drawn with dashed lines of the same color.}
    \label{fig:sigmoid}
\end{figure}
Fig.~\ref{fig:sigmoid} shows results for this difference obtained from our numerical solution in the two-time plane for $q = 4$.
One can see that the closer one stays to the time diagonal (small $\tau$), the more step-like the change of the Green's function becomes.
For further analysis, we use a sigmoid function ansatz over $t_\text{avg} = t_1-\tau/2$ to describe the time ordered Green's functions' \eqref{time_orderedG} differences as a function of $t_1$,
\begin{align}
    \vert \Gg(t_1,t_1-\tau) - \Gg_{\text{eq}}(\tau) \vert \propto \frac{1}{1+ e^{k t_\text{avg}}} .
    \label{sigmoid}
\end{align}
The slope parameter $k$ is extracted from a fit to the data and shown in Fig.~\ref{kfit}.
At small $\tau$, but non-zero $\tau$, the data is consistent with a near-universal rate 
 \begin{equation}
 k = \frac{2\pi}{\tau} [1+\Oo(10^{-1}) ] ,\label{k_rate}
 \end{equation}
where the non-universal corrections in $\Oo(10^{-1})$ are an order of magnitude smaller. Note that strictly in the limit as $\tau \to 0$, we are left subtracting two boundary terms from another $\Gg(t_1,t_1) = -\imath/2$ and $\Gg_{\text{eq}}(0)= -\imath/2$. As such any sigmoid fit in this limit is meaningless. This is further seen by observing that the time arguments $(t_1,t_2= t_1-\tau)$ lie in the two-time plane block $(1,0)$ (cf. Fig.~\ref{fig:blocks}) on a line of width $\tau$. The rate \eqref{k_rate} is of a different nature to the one considered in Sec.~\ref{secTeff}. It is inversely proportional to $\tau$ at which the correlator is probed and seemingly only weakly dependent on $q$ and $T$.

\section{Conclusion}
\label{conclusion section} In this work we have tested the notion of ``instantaneous thermalization'' by considering quenches of SYK models involving only $q/2$-body interactions. We solve the problem numerically for a range of finite $q$ values. We compared this to the recent analytical solution, which solves for the entire time plane to leading order in $1/q$ \cite{CompanionLouw2026}. The latter shows instantaneous post-quench thermalization to leading order. Seemingly paradoxically, we saw a decaying thermalization rate $\Gamma \propto 1/q$. We confirmed that the analytical results from \cite{CompanionLouw2026} faithfully represent the numerical correlation for all times. This leads one to ask: what is really meant by ``instantaneous thermalization''?

Our findings provide insights via a time-block-resolved picture.
While the diagonal time-block correlations, and thus the post-quench (those in block $(1,1)$) Green's functions, remain thermal, any measure requiring information over the connecting ($(1,0)$ or $(0,1)$) blocks inherits a finite rate from inherently non-thermal off-diagonal block correlations.
This means that there will always be contributions that require a finite time to relax.
Hence, the effective temperature does not change instantaneously during the quench. We made this quantitative in Sec.~\ref{secblocks}, where the off-diagonal Green's functions were found to cross over on the line $t_\text{avg}=0$ with a width $\tau/2\pi$ set by the relative time alone, independently of both $T$ and $q$.

The three measures we have considered therefore return three different answers: a vanishing difference to the large-$q$ theory in Sec.~\ref{secGdiff}, a Planckian $\Gamma \sim T q^{-1}$ in Sec.~\ref{secTeff}, and a probe-dependent $k$ trending as $2\pi/\tau$ in Sec.~\ref{secblocks}.
We have studied effective temperatures for the non-equilibrium Green's function in the standard Wigner coordinates, following the work by Eberlein~\cite{Eberlein2017Jun}.
However, if one chooses to perform the Fourier transform to frequency space with respect to a different choice of coordinates (e.g. only forward in time for $\tau$), one can also expect different effective temperatures and rates.
A closer inspection of possible Fourier transforms (and their physical relevance) could be an interesting direction of further study.

In summary, the question of our title therefore admits no unique answer: the rate one obtains is set by the notion of thermalization one adopts.
Nevertheless, our analysis helps to bridge the gap between finite and infinite-$q$ dynamics.
Future directions could consider whether these results can be extended to other quench protocols or transport applications. Our model also opens the door to further analytical investigations of the out-of-equilibrium dynamics, especially from quenching local operators in local generalizations of our model \cite{Song2017Nov}.
Lastly, assuming the thermalization rates in Fig.~\ref{fig:gammas_vs_qs} go to zero as $q\to\infty$, what would this imply in terms of capturing thermalization? Does it imply a diverging memory effect? To address this, one could solve for the effective temperature analytically. Then one could also assess if (and in what sense) the $1/q$ scaling is a trivial result.

\begin{acknowledgments}
A.O. and J.C.L. acknowledge helpful discussions with Stefan Kehrein. J.C.L.\ was employed at the Max Planck Institute of Quantum Optics and, held a teaching appointment at the Faculty of Physics of Ludwig Maximilian University of Munich during the main course of this work. J.C.L. acknowledges funding by the Max Planck Society and by the Deutsche Forschungsgemeinschaft (DFG, German Research Foundation) under Germany's Excellence Strategy, EXC-2111 - 390814868. A.O. was employed at Jo\v{z}ef Stefan Institute during the main course of this work.
A.O. acknowledges support from No. P1-0044 of the Slovenian Research Agency (ARIS).
\end{acknowledgments}

\bibliography{ref.bib}

\begin{thebibliography}{30}%
\makeatletter
\providecommand \@ifxundefined [1]{%
 \@ifx{#1\undefined}
}%
\providecommand \@ifnum [1]{%
 \ifnum #1\expandafter \@firstoftwo
 \else \expandafter \@secondoftwo
 \fi
}%
\providecommand \@ifx [1]{%
 \ifx #1\expandafter \@firstoftwo
 \else \expandafter \@secondoftwo
 \fi
}%
\providecommand \natexlab [1]{#1}%
\providecommand \enquote  [1]{``#1''}%
\providecommand \bibnamefont  [1]{#1}%
\providecommand \bibfnamefont [1]{#1}%
\providecommand \citenamefont [1]{#1}%
\providecommand \href@noop [0]{\@secondoftwo}%
\providecommand \href [0]{\begingroup \@sanitize@url \@href}%
\providecommand \@href[1]{\@@startlink{#1}\@@href}%
\providecommand \@@href[1]{\endgroup#1\@@endlink}%
\providecommand \@sanitize@url [0]{\catcode `\\12\catcode `\$12\catcode
  `\&12\catcode `\#12\catcode `\^12\catcode `\_12\catcode `\%12\relax}%
\providecommand \@@startlink[1]{}%
\providecommand \@@endlink[0]{}%
\providecommand \url  [0]{\begingroup\@sanitize@url \@url }%
\providecommand \@url [1]{\endgroup\@href {#1}{\urlprefix }}%
\providecommand \urlprefix  [0]{URL }%
\providecommand \Eprint [0]{\href }%
\providecommand \doibase [0]{https://doi.org/}%
\providecommand \selectlanguage [0]{\@gobble}%
\providecommand \bibinfo  [0]{\@secondoftwo}%
\providecommand \bibfield  [0]{\@secondoftwo}%
\providecommand \translation [1]{[#1]}%
\providecommand \BibitemOpen [0]{}%
\providecommand \bibitemStop [0]{}%
\providecommand \bibitemNoStop [0]{.\EOS\space}%
\providecommand \EOS [0]{\spacefactor3000\relax}%
\providecommand \BibitemShut  [1]{\csname bibitem#1\endcsname}%
\let\auto@bib@innerbib\@empty
\bibitem [{\citenamefont {Sachdev}\ and\ \citenamefont
  {Ye}(1993)}]{SachdevYe1992}%
  \BibitemOpen
  \bibfield  {author} {\bibinfo {author} {\bibfnamefont {S.}~\bibnamefont
  {Sachdev}}\ and\ \bibinfo {author} {\bibfnamefont {J.}~\bibnamefont {Ye}},\
  }\bibfield  {title} {\bibinfo {title} {{Gapless spin-fluid ground state in a
  random quantum Heisenberg magnet}},\ }\href
  {https://doi.org/10.1103/PhysRevLett.70.3339} {\bibfield  {journal} {\bibinfo
   {journal} {Phys. Rev. Lett.}\ }\textbf {\bibinfo {volume} {70}},\ \bibinfo
  {pages} {3339} (\bibinfo {year} {1993})}\BibitemShut {NoStop}%
\bibitem [{\citenamefont {Kitaev}(2015)}]{Kitaev2015}%
  \BibitemOpen
  \bibfield  {author} {\bibinfo {author} {\bibfnamefont {A.}~\bibnamefont
  {Kitaev}},\ }\href@noop {} {\bibinfo {title} {{A simple model of quantum
  holography}}},\ \bibinfo {howpublished} {Talks at the KITP program {\it
  Entanglement in Strongly-Correlated Quantum Matter},
  \href{https://online.kitp.ucsb.edu/online/entangled15/kitaev/}{(part 1)} and
  \href{https://online.kitp.ucsb.edu/online/entangled15/kitaev2/}{(part 2)}}
  (\bibinfo {year} {2015})\BibitemShut {NoStop}%
\bibitem [{\citenamefont {Chowdhury}\ \emph {et~al.}(2022)\citenamefont
  {Chowdhury}, \citenamefont {Georges}, \citenamefont {Parcollet},\ and\
  \citenamefont {Sachdev}}]{Chowdhury2022Sep}%
  \BibitemOpen
  \bibfield  {author} {\bibinfo {author} {\bibfnamefont {D.}~\bibnamefont
  {Chowdhury}}, \bibinfo {author} {\bibfnamefont {A.}~\bibnamefont {Georges}},
  \bibinfo {author} {\bibfnamefont {O.}~\bibnamefont {Parcollet}},\ and\
  \bibinfo {author} {\bibfnamefont {S.}~\bibnamefont {Sachdev}},\ }\bibfield
  {title} {\bibinfo {title} {{Sachdev-Ye-Kitaev models and beyond: Window into
  non-Fermi liquids}},\ }\href {https://doi.org/10.1103/RevModPhys.94.035004}
  {\bibfield  {journal} {\bibinfo  {journal} {Rev. Mod. Phys.}\ }\textbf
  {\bibinfo {volume} {94}},\ \bibinfo {pages} {035004} (\bibinfo {year}
  {2022})}\BibitemShut {NoStop}%
\bibitem [{\citenamefont {Eberlein}\ \emph {et~al.}(2017)\citenamefont
  {Eberlein}, \citenamefont {Kasper}, \citenamefont {Sachdev},\ and\
  \citenamefont {Steinberg}}]{Eberlein2017Jun}%
  \BibitemOpen
  \bibfield  {author} {\bibinfo {author} {\bibfnamefont {A.}~\bibnamefont
  {Eberlein}}, \bibinfo {author} {\bibfnamefont {V.}~\bibnamefont {Kasper}},
  \bibinfo {author} {\bibfnamefont {S.}~\bibnamefont {Sachdev}},\ and\ \bibinfo
  {author} {\bibfnamefont {J.}~\bibnamefont {Steinberg}},\ }\bibfield  {title}
  {\bibinfo {title} {{Quantum quench of the Sachdev-Ye-Kitaev model}},\ }\href
  {https://doi.org/10.1103/PhysRevB.96.205123} {\bibfield  {journal} {\bibinfo
  {journal} {Phys. Rev. B}\ }\textbf {\bibinfo {volume} {96}},\ \bibinfo
  {pages} {205123} (\bibinfo {year} {2017})}\BibitemShut {NoStop}%
\bibitem [{\citenamefont {Louw}\ and\ \citenamefont
  {Kehrein}(2022)}]{Louw2022Feb}%
  \BibitemOpen
  \bibfield  {author} {\bibinfo {author} {\bibfnamefont {J.~C.}\ \bibnamefont
  {Louw}}\ and\ \bibinfo {author} {\bibfnamefont {S.}~\bibnamefont {Kehrein}},\
  }\bibfield  {title} {\bibinfo {title} {{Thermalization of many many-body
  interacting Sachdev-Ye-Kitaev models}},\ }\href
  {https://doi.org/10.1103/PhysRevB.105.075117} {\bibfield  {journal} {\bibinfo
   {journal} {Phys. Rev. B}\ }\textbf {\bibinfo {volume} {105}},\ \bibinfo
  {pages} {075117} (\bibinfo {year} {2022})}\BibitemShut {NoStop}%
\bibitem [{\citenamefont {Almheiri}\ \emph {et~al.}(2024)\citenamefont
  {Almheiri}, \citenamefont {Milekhin},\ and\ \citenamefont
  {Swingle}}]{Almheiri2024Aug}%
  \BibitemOpen
  \bibfield  {author} {\bibinfo {author} {\bibfnamefont {A.}~\bibnamefont
  {Almheiri}}, \bibinfo {author} {\bibfnamefont {A.}~\bibnamefont {Milekhin}},\
  and\ \bibinfo {author} {\bibfnamefont {B.}~\bibnamefont {Swingle}},\
  }\bibfield  {title} {\bibinfo {title} {{Universal constraints on energy flow
  and SYK thermalization}},\ }\href {https://doi.org/10.1007/JHEP08(2024)034}
  {\bibfield  {journal} {\bibinfo  {journal} {J. High Energy Phys.}\ }\textbf
  {\bibinfo {volume} {2024}}\bibinfo  {number} { (8)},\ \bibinfo {pages}
  {34}}\BibitemShut {NoStop}%
\bibitem [{\citenamefont {Sohal}\ \emph {et~al.}(2022)\citenamefont {Sohal},
  \citenamefont {Nie}, \citenamefont {Sun},\ and\ \citenamefont
  {Fradkin}}]{Sohal2022Jan}%
  \BibitemOpen
\bibfield  {number} {  }\bibfield  {author} {\bibinfo {author} {\bibfnamefont
  {R.}~\bibnamefont {Sohal}}, \bibinfo {author} {\bibfnamefont
  {L.}~\bibnamefont {Nie}}, \bibinfo {author} {\bibfnamefont {X.-Q.}\
  \bibnamefont {Sun}},\ and\ \bibinfo {author} {\bibfnamefont {E.}~\bibnamefont
  {Fradkin}},\ }\bibfield  {title} {\bibinfo {title} {{Thermalization of
  randomly coupled SYK models}},\ }\href
  {https://doi.org/10.1088/1742-5468/ac416b} {\bibfield  {journal} {\bibinfo
  {journal} {J. Stat. Mech.: Theory Exp.}\ }\textbf {\bibinfo {volume}
  {2022}}\bibinfo  {number} { (1)},\ \bibinfo {pages} {013103}}\BibitemShut
  {NoStop}%
\bibitem [{\citenamefont {Halataei}(2025)}]{Halataei2025Jul}%
  \BibitemOpen
\bibfield  {number} {  }\bibfield  {author} {\bibinfo {author} {\bibfnamefont
  {S.~M.~H.}\ \bibnamefont {Halataei}},\ }\bibfield  {title} {\bibinfo {title}
  {{Eigenstate thermalization in the two-site SYK and SYK chain models}},\
  }\href {https://doi.org/10.1103/33fm-syj6} {\bibfield  {journal} {\bibinfo
  {journal} {Phys. Rev. D}\ }\textbf {\bibinfo {volume} {112}},\ \bibinfo
  {pages} {026016} (\bibinfo {year} {2025})}\BibitemShut {NoStop}%
\bibitem [{\citenamefont {Jaramillo}\ \emph {et~al.}(2025)\citenamefont
  {Jaramillo}, \citenamefont {Jha},\ and\ \citenamefont
  {Kehrein}}]{Jaramillo2025May}%
  \BibitemOpen
  \bibfield  {author} {\bibinfo {author} {\bibfnamefont {S.~S.}\ \bibnamefont
  {Jaramillo}}, \bibinfo {author} {\bibfnamefont {R.}~\bibnamefont {Jha}},\
  and\ \bibinfo {author} {\bibfnamefont {S.}~\bibnamefont {Kehrein}},\
  }\bibfield  {title} {\bibinfo {title} {{Thermalization of a closed
  Sachdev-Ye-Kitaev system in the thermodynamic limit}},\ }\href
  {https://doi.org/10.1103/PhysRevB.111.195153} {\bibfield  {journal} {\bibinfo
   {journal} {Phys. Rev. B}\ }\textbf {\bibinfo {volume} {111}},\ \bibinfo
  {pages} {195153} (\bibinfo {year} {2025})}\BibitemShut {NoStop}%
\bibitem [{\citenamefont {Perugu}\ \emph {et~al.}(2025)\citenamefont {Perugu},
  \citenamefont {Haldar},\ and\ \citenamefont {Banerjee}}]{Perugu2025Nov}%
  \BibitemOpen
  \bibfield  {author} {\bibinfo {author} {\bibfnamefont {R.}~\bibnamefont
  {Perugu}}, \bibinfo {author} {\bibfnamefont {A.}~\bibnamefont {Haldar}},\
  and\ \bibinfo {author} {\bibfnamefont {S.}~\bibnamefont {Banerjee}},\
  }\bibfield  {title} {\bibinfo {title} {{Universal nonequilibrium dynamics of
  pure states and density-dependent thermalization in the Sachdev-Ye-Kitaev
  model}},\ }\href {https://doi.org/10.1103/9cbp-7tvv} {\bibfield  {journal}
  {\bibinfo  {journal} {Phys. Rev. B}\ }\textbf {\bibinfo {volume} {112}},\
  \bibinfo {pages} {184301} (\bibinfo {year} {2025})}\BibitemShut {NoStop}%
\bibitem [{\citenamefont {Bhattacharya}\ \emph {et~al.}(2019)\citenamefont
  {Bhattacharya}, \citenamefont {Jatkar},\ and\ \citenamefont
  {Sorokhaibam}}]{Bhattacharya2019Jul}%
  \BibitemOpen
  \bibfield  {author} {\bibinfo {author} {\bibfnamefont {R.}~\bibnamefont
  {Bhattacharya}}, \bibinfo {author} {\bibfnamefont {D.~P.}\ \bibnamefont
  {Jatkar}},\ and\ \bibinfo {author} {\bibfnamefont {N.}~\bibnamefont
  {Sorokhaibam}},\ }\bibfield  {title} {\bibinfo {title} {{Quantum quenches and
  thermalization in SYK models}},\ }\href
  {https://doi.org/10.1007/JHEP07(2019)066} {\bibfield  {journal} {\bibinfo
  {journal} {J. High Energy Phys.}\ }\textbf {\bibinfo {volume} {2019}}\bibinfo
   {number} { (7)},\ \bibinfo {pages} {66}}\BibitemShut {NoStop}%
\bibitem [{\citenamefont {Bandyopadhyay}\ \emph {et~al.}(2023)\citenamefont
  {Bandyopadhyay}, \citenamefont {Uhrich}, \citenamefont {Paviglianiti},\ and\
  \citenamefont {Hauke}}]{Bandyopadhyay2023May}%
  \BibitemOpen
\bibfield  {number} {  }\bibfield  {author} {\bibinfo {author} {\bibfnamefont
  {S.}~\bibnamefont {Bandyopadhyay}}, \bibinfo {author} {\bibfnamefont
  {P.}~\bibnamefont {Uhrich}}, \bibinfo {author} {\bibfnamefont
  {A.}~\bibnamefont {Paviglianiti}},\ and\ \bibinfo {author} {\bibfnamefont
  {P.}~\bibnamefont {Hauke}},\ }\bibfield  {title} {\bibinfo {title}
  {{Universal equilibration dynamics of the Sachdev-Ye-Kitaev model}},\ }\href
  {https://doi.org/10.22331/q-2023-05-24-1022} {\bibfield  {journal} {\bibinfo
  {journal} {Quantum}\ }\textbf {\bibinfo {volume} {7}},\ \bibinfo {pages}
  {1022} (\bibinfo {year} {2023})}\BibitemShut {NoStop}%
\bibitem [{\citenamefont {Azeyanagi}\ \emph {et~al.}(2018)\citenamefont
  {Azeyanagi}, \citenamefont {Ferrari},\ and\ \citenamefont
  {Massolo}}]{Azeyanagi2018Feb}%
  \BibitemOpen
  \bibfield  {author} {\bibinfo {author} {\bibfnamefont {T.}~\bibnamefont
  {Azeyanagi}}, \bibinfo {author} {\bibfnamefont {F.}~\bibnamefont {Ferrari}},\
  and\ \bibinfo {author} {\bibfnamefont {F.~I.~S.}\ \bibnamefont {Massolo}},\
  }\bibfield  {title} {\bibinfo {title} {{Phase Diagram of Planar Matrix
  Quantum Mechanics, Tensor, and Sachdev-Ye-Kitaev Models}},\ }\href
  {https://doi.org/10.1103/PhysRevLett.120.061602} {\bibfield  {journal}
  {\bibinfo  {journal} {Phys. Rev. Lett.}\ }\textbf {\bibinfo {volume} {120}},\
  \bibinfo {pages} {061602} (\bibinfo {year} {2018})}\BibitemShut {NoStop}%
\bibitem [{\citenamefont {Ferrari}\ and\ \citenamefont
  {Schaposnik~Massolo}(2019)}]{Ferrari2019Jul}%
  \BibitemOpen
  \bibfield  {author} {\bibinfo {author} {\bibfnamefont {F.}~\bibnamefont
  {Ferrari}}\ and\ \bibinfo {author} {\bibfnamefont {F.~I.}\ \bibnamefont
  {Schaposnik~Massolo}},\ }\bibfield  {title} {\bibinfo {title} {{Phases of
  melonic quantum mechanics}},\ }\href
  {https://doi.org/10.1103/PhysRevD.100.026007} {\bibfield  {journal} {\bibinfo
   {journal} {Phys. Rev. D}\ }\textbf {\bibinfo {volume} {100}},\ \bibinfo
  {pages} {026007} (\bibinfo {year} {2019})}\BibitemShut {NoStop}%
\bibitem [{\citenamefont {Louw}\ and\ \citenamefont
  {Kehrein}(2023)}]{Louw2023Feb}%
  \BibitemOpen
  \bibfield  {author} {\bibinfo {author} {\bibfnamefont {J.~C.}\ \bibnamefont
  {Louw}}\ and\ \bibinfo {author} {\bibfnamefont {S.}~\bibnamefont {Kehrein}},\
  }\bibfield  {title} {\bibinfo {title} {{Shared universality of charged black
  holes and the complex large-$q$ Sachdev-Ye-Kitaev model}},\ }\href
  {https://doi.org/10.1103/PhysRevB.107.075132} {\bibfield  {journal} {\bibinfo
   {journal} {Phys. Rev. B}\ }\textbf {\bibinfo {volume} {107}},\ \bibinfo
  {pages} {075132} (\bibinfo {year} {2023})}\BibitemShut {NoStop}%
\bibitem [{\citenamefont {Zanoci}\ and\ \citenamefont
  {Swingle}(2022)}]{Zanoci2022Apr}%
  \BibitemOpen
  \bibfield  {author} {\bibinfo {author} {\bibfnamefont {C.}~\bibnamefont
  {Zanoci}}\ and\ \bibinfo {author} {\bibfnamefont {B.}~\bibnamefont
  {Swingle}},\ }\bibfield  {title} {\bibinfo {title} {{Energy transport in
  Sachdev-Ye-Kitaev networks coupled to thermal baths}},\ }\href
  {https://doi.org/10.1103/PhysRevResearch.4.023001} {\bibfield  {journal}
  {\bibinfo  {journal} {Phys. Rev. Res.}\ }\textbf {\bibinfo {volume} {4}},\
  \bibinfo {pages} {023001} (\bibinfo {year} {2022})}\BibitemShut {NoStop}%
\bibitem [{\citenamefont {Jha}\ \emph {et~al.}(2025)\citenamefont {Jha},
  \citenamefont {Kehrein},\ and\ \citenamefont {Louw}}]{Jha2025Jan}%
  \BibitemOpen
  \bibfield  {author} {\bibinfo {author} {\bibfnamefont {R.}~\bibnamefont
  {Jha}}, \bibinfo {author} {\bibfnamefont {S.}~\bibnamefont {Kehrein}},\ and\
  \bibinfo {author} {\bibfnamefont {J.~C.}\ \bibnamefont {Louw}},\ }\bibfield
  {title} {\bibinfo {title} {{Current correlations and conductivity in SYK-like
  systems: An analytical study}},\ }\href
  {https://doi.org/10.1103/PhysRevB.111.045111} {\bibfield  {journal} {\bibinfo
   {journal} {Phys. Rev. B}\ }\textbf {\bibinfo {volume} {111}},\ \bibinfo
  {pages} {045111} (\bibinfo {year} {2025})}\BibitemShut {NoStop}%
\bibitem [{\citenamefont {Maldacena}\ \emph {et~al.}(2016)\citenamefont
  {Maldacena}, \citenamefont {Shenker},\ and\ \citenamefont
  {Stanford}}]{Maldacena2016Aug}%
  \BibitemOpen
  \bibfield  {author} {\bibinfo {author} {\bibfnamefont {J.}~\bibnamefont
  {Maldacena}}, \bibinfo {author} {\bibfnamefont {S.~H.}\ \bibnamefont
  {Shenker}},\ and\ \bibinfo {author} {\bibfnamefont {D.}~\bibnamefont
  {Stanford}},\ }\bibfield  {title} {\bibinfo {title} {{A bound on chaos}},\
  }\href {https://doi.org/10.1007/JHEP08(2016)106} {\bibfield  {journal}
  {\bibinfo  {journal} {J. High Energy Phys.}\ }\textbf {\bibinfo {volume}
  {2016}}\bibinfo  {number} { (8)},\ \bibinfo {pages} {1}}\BibitemShut
  {NoStop}%
\bibitem [{\citenamefont {Lei}\ and\ \citenamefont {Ge}(2022)}]{Lei2022Apr}%
  \BibitemOpen
\bibfield  {number} {  }\bibfield  {author} {\bibinfo {author} {\bibfnamefont
  {Y.-Q.}\ \bibnamefont {Lei}}\ and\ \bibinfo {author} {\bibfnamefont {X.-H.}\
  \bibnamefont {Ge}},\ }\bibfield  {title} {\bibinfo {title} {{Circular motion
  of charged particles near a charged black hole}},\ }\href
  {https://doi.org/10.1103/PhysRevD.105.084011} {\bibfield  {journal} {\bibinfo
   {journal} {Phys. Rev. D}\ }\textbf {\bibinfo {volume} {105}},\ \bibinfo
  {pages} {084011} (\bibinfo {year} {2022})}\BibitemShut {NoStop}%
\bibitem [{\citenamefont {Vishveshwara}(1970)}]{Vishveshwara1970Aug}%
  \BibitemOpen
  \bibfield  {author} {\bibinfo {author} {\bibfnamefont {C.~V.}\ \bibnamefont
  {Vishveshwara}},\ }\bibfield  {title} {\bibinfo {title} {{Scattering of
  Gravitational Radiation by a Schwarzschild Black-hole}},\ }\href
  {https://doi.org/10.1038/227936a0} {\bibfield  {journal} {\bibinfo  {journal}
  {Nature}\ }\textbf {\bibinfo {volume} {227}},\ \bibinfo {pages} {936}
  (\bibinfo {year} {1970})}\BibitemShut {NoStop}%
\bibitem [{\citenamefont {Carullo}\ \emph {et~al.}(2021)\citenamefont
  {Carullo}, \citenamefont {Laghi}, \citenamefont {Veitch},\ and\ \citenamefont
  {Del~Pozzo}}]{Carullo2021Apr}%
  \BibitemOpen
  \bibfield  {author} {\bibinfo {author} {\bibfnamefont {G.}~\bibnamefont
  {Carullo}}, \bibinfo {author} {\bibfnamefont {D.}~\bibnamefont {Laghi}},
  \bibinfo {author} {\bibfnamefont {J.}~\bibnamefont {Veitch}},\ and\ \bibinfo
  {author} {\bibfnamefont {W.}~\bibnamefont {Del~Pozzo}},\ }\bibfield  {title}
  {\bibinfo {title} {{Bekenstein-Hod Universal Bound on Information Emission
  Rate Is Obeyed by LIGO-Virgo Binary Black Hole Remnants}},\ }\href
  {https://doi.org/10.1103/PhysRevLett.126.161102} {\bibfield  {journal}
  {\bibinfo  {journal} {Phys. Rev. Lett.}\ }\textbf {\bibinfo {volume} {126}},\
  \bibinfo {pages} {161102} (\bibinfo {year} {2021})}\BibitemShut {NoStop}%
\bibitem [{\citenamefont {Louw}\ \emph {et~al.}(2023)\citenamefont {Louw},
  \citenamefont {Cao},\ and\ \citenamefont {Ge}}]{Louw2023Oct}%
  \BibitemOpen
  \bibfield  {author} {\bibinfo {author} {\bibfnamefont {J.~C.}\ \bibnamefont
  {Louw}}, \bibinfo {author} {\bibfnamefont {S.}~\bibnamefont {Cao}},\ and\
  \bibinfo {author} {\bibfnamefont {X.-H.}\ \bibnamefont {Ge}},\ }\bibfield
  {title} {\bibinfo {title} {{Matching partition functions of deformed
  Jackiw-Teitelboim gravity and the complex SYK model}},\ }\href
  {https://doi.org/10.1103/PhysRevD.108.086014} {\bibfield  {journal} {\bibinfo
   {journal} {Phys. Rev. D}\ }\textbf {\bibinfo {volume} {108}},\ \bibinfo
  {pages} {086014} (\bibinfo {year} {2023})}\BibitemShut {NoStop}%
\bibitem [{\citenamefont {Goldstein}\ \emph {et~al.}(2015)\citenamefont
  {Goldstein}, \citenamefont {Hara},\ and\ \citenamefont
  {Tasaki}}]{Goldstein2015Apr}%
  \BibitemOpen
  \bibfield  {author} {\bibinfo {author} {\bibfnamefont {S.}~\bibnamefont
  {Goldstein}}, \bibinfo {author} {\bibfnamefont {T.}~\bibnamefont {Hara}},\
  and\ \bibinfo {author} {\bibfnamefont {H.}~\bibnamefont {Tasaki}},\
  }\bibfield  {title} {\bibinfo {title} {{Extremely quick thermalization in a
  macroscopic quantum system for a typical nonequilibrium subspace}},\ }\href
  {https://doi.org/10.1088/1367-2630/17/4/045002} {\bibfield  {journal}
  {\bibinfo  {journal} {New J. Phys.}\ }\textbf {\bibinfo {volume} {17}},\
  \bibinfo {pages} {045002} (\bibinfo {year} {2015})}\BibitemShut {NoStop}%
\bibitem [{\citenamefont {Nickelsen}\ and\ \citenamefont
  {Kastner}(2019)}]{Nickelsen2019May}%
  \BibitemOpen
  \bibfield  {author} {\bibinfo {author} {\bibfnamefont {D.}~\bibnamefont
  {Nickelsen}}\ and\ \bibinfo {author} {\bibfnamefont {M.}~\bibnamefont
  {Kastner}},\ }\bibfield  {title} {\bibinfo {title} {{Classical Lieb-Robinson
  Bound for Estimating Equilibration Timescales of Isolated Quantum Systems}},\
  }\href {https://doi.org/10.1103/PhysRevLett.122.180602} {\bibfield  {journal}
  {\bibinfo  {journal} {Phys. Rev. Lett.}\ }\textbf {\bibinfo {volume} {122}},\
  \bibinfo {pages} {180602} (\bibinfo {year} {2019})}\BibitemShut {NoStop}%
\bibitem [{\citenamefont {Pappalardi}\ \emph {et~al.}(2022)\citenamefont
  {Pappalardi}, \citenamefont {Foini},\ and\ \citenamefont
  {Kurchan}}]{Pappalardi2022Apr}%
  \BibitemOpen
  \bibfield  {author} {\bibinfo {author} {\bibfnamefont {S.}~\bibnamefont
  {Pappalardi}}, \bibinfo {author} {\bibfnamefont {L.}~\bibnamefont {Foini}},\
  and\ \bibinfo {author} {\bibfnamefont {J.}~\bibnamefont {Kurchan}},\
  }\bibfield  {title} {\bibinfo {title} {{Quantum bounds and
  fluctuation-dissipation relations}},\ }\href
  {https://doi.org/10.21468/SciPostPhys.12.4.130} {\bibfield  {journal}
  {\bibinfo  {journal} {SciPost Phys.}\ }\textbf {\bibinfo {volume} {12}},\
  \bibinfo {pages} {130} (\bibinfo {year} {2022})}\BibitemShut {NoStop}%
\bibitem [{\citenamefont {Louw}(2026)}]{CompanionLouw2026}%
  \BibitemOpen
  \bibfield  {author} {\bibinfo {author} {\bibfnamefont {J.~C.}\ \bibnamefont
  {Louw}},\ }\bibfield  {title} {\bibinfo {title} {{Analytical solutions to the
  non-equilibrium Green's functions in large-$q$ SYK models}}} (\bibinfo {year}
  {2026}),\ \bibinfo {note} {companion paper, posted on arXiv concurrently with
  the present work}\BibitemShut {NoStop}%
\bibitem [{\citenamefont {Jha}\ and\ \citenamefont {Louw}(2023)}]{Jha2023Jun}%
  \BibitemOpen
  \bibfield  {author} {\bibinfo {author} {\bibfnamefont {R.}~\bibnamefont
  {Jha}}\ and\ \bibinfo {author} {\bibfnamefont {J.~C.}\ \bibnamefont {Louw}},\
  }\bibfield  {title} {\bibinfo {title} {{Dynamics and charge fluctuations in
  large-$q$ Sachdev-Ye-Kitaev lattices}},\ }\href
  {https://doi.org/10.1103/PhysRevB.107.235114} {\bibfield  {journal} {\bibinfo
   {journal} {Phys. Rev. B}\ }\textbf {\bibinfo {volume} {107}},\ \bibinfo
  {pages} {235114} (\bibinfo {year} {2023})}\BibitemShut {NoStop}%
\bibitem [{\citenamefont {Sch\"uler}\ \emph {et~al.}(2020)\citenamefont
  {Sch\"uler}, \citenamefont {Gole{\v{z}}}, \citenamefont {Murakami},
  \citenamefont {Bittner}, \citenamefont {Herrmann}, \citenamefont {Strand},
  \citenamefont {Werner},\ and\ \citenamefont {Eckstein}}]{Schueler2020}%
  \BibitemOpen
  \bibfield  {author} {\bibinfo {author} {\bibfnamefont {M.}~\bibnamefont
  {Sch\"uler}}, \bibinfo {author} {\bibfnamefont {D.}~\bibnamefont
  {Gole{\v{z}}}}, \bibinfo {author} {\bibfnamefont {Y.}~\bibnamefont
  {Murakami}}, \bibinfo {author} {\bibfnamefont {N.}~\bibnamefont {Bittner}},
  \bibinfo {author} {\bibfnamefont {A.}~\bibnamefont {Herrmann}}, \bibinfo
  {author} {\bibfnamefont {H.~U.}\ \bibnamefont {Strand}}, \bibinfo {author}
  {\bibfnamefont {P.}~\bibnamefont {Werner}},\ and\ \bibinfo {author}
  {\bibfnamefont {M.}~\bibnamefont {Eckstein}},\ }\bibfield  {title} {\bibinfo
  {title} {Nessi: The non-equilibrium systems simulation package},\ }\href
  {https://doi.org/10.1016/j.cpc.2020.107484} {\bibfield  {journal} {\bibinfo
  {journal} {Comp. Phys Comm.}\ }\textbf {\bibinfo {volume} {257}},\ \bibinfo
  {pages} {107484} (\bibinfo {year} {2020})}\BibitemShut {NoStop}%
\bibitem [{\citenamefont {Song}\ \emph {et~al.}(2017)\citenamefont {Song},
  \citenamefont {Jian},\ and\ \citenamefont {Balents}}]{Song2017Nov}%
  \BibitemOpen
  \bibfield  {author} {\bibinfo {author} {\bibfnamefont {X.-Y.}\ \bibnamefont
  {Song}}, \bibinfo {author} {\bibfnamefont {C.-M.}\ \bibnamefont {Jian}},\
  and\ \bibinfo {author} {\bibfnamefont {L.}~\bibnamefont {Balents}},\
  }\bibfield  {title} {\bibinfo {title} {{Strongly Correlated Metal Built from
  Sachdev-Ye-Kitaev Models}},\ }\href
  {https://doi.org/10.1103/PhysRevLett.119.216601} {\bibfield  {journal}
  {\bibinfo  {journal} {Phys. Rev. Lett.}\ }\textbf {\bibinfo {volume} {119}},\
  \bibinfo {pages} {216601} (\bibinfo {year} {2017})}\BibitemShut {NoStop}%
\bibitem [{\citenamefont {Maldacena}\ and\ \citenamefont
  {Stanford}(2016)}]{Maldacena2016Nov}%
  \BibitemOpen
  \bibfield  {author} {\bibinfo {author} {\bibfnamefont {J.}~\bibnamefont
  {Maldacena}}\ and\ \bibinfo {author} {\bibfnamefont {D.}~\bibnamefont
  {Stanford}},\ }\bibfield  {title} {\bibinfo {title} {{Remarks on the
  Sachdev-Ye-Kitaev model}},\ }\href
  {https://doi.org/10.1103/PhysRevD.94.106002} {\bibfield  {journal} {\bibinfo
  {journal} {Phys. Rev. D}\ }\textbf {\bibinfo {volume} {94}},\ \bibinfo
  {pages} {106002} (\bibinfo {year} {2016})}\BibitemShut {NoStop}%
\end{thebibliography}%

\appendix

\section{Numerical extraction of effective temperatures\label{app:numerics}}
\begin{figure}
\includegraphics[width=0.35\textwidth]{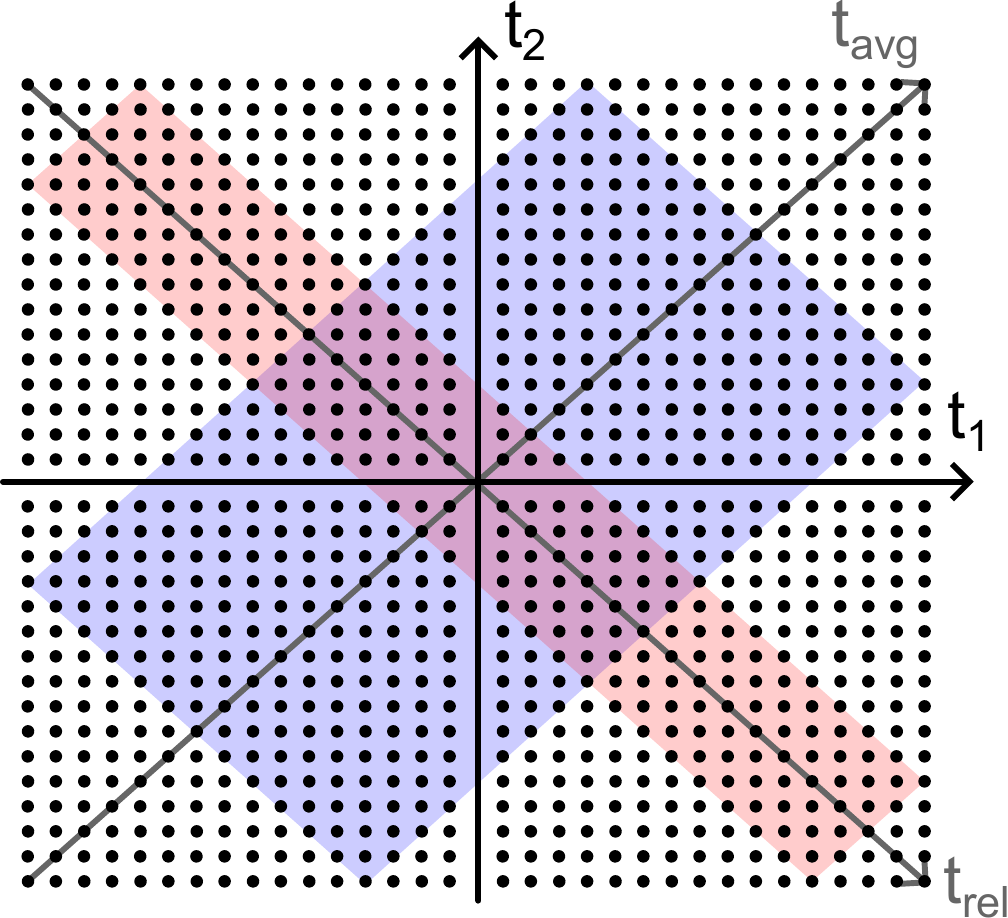}
\caption{Sketch of the discrete two-time blocks used for the calculations in this paper. We exemplarily show two tilted rectangles that could be used for performing the Fourier transform with respect to $\tau = t_\text{rel}$.\label{appfig:timeplane_bketch}}
\end{figure}

In the following, we will provide more information on our data processing and a discussion of how the thermalization rates were extracted from the Green's function data generated via NESSi \cite{Schueler2020}.
The first step is to Fourier-transform the real-time Green's function with respect to the relative time $\tau = t_\text{rel} := t_1 - t_2$.
As the remaining real time argument we choose the orthogonal average time $t_\text{avg} = ( t_1 + t_2 ) / 2$, giving rise to Wigner coordinates.
Our numerical solutions for the Green's function $G(t_1, t_2)$ are obtained for a fixed rectangle in the $t_1$-$t_2$ plane such that one needs to define a tilted rectangle, which is used for a consistent transformation to time-frequency coordinates $(t_\text{avg}, \omega)$.
Such a selection is sketched in Fig.~\ref{appfig:timeplane_bketch}.
Since effective temperatures $T_\text{eff}$ are fitted from the low-frequency behavior of the Green's functions, one needs to make sure that sufficiently long relative times are available for the Fourier transform, while still being able to access the relevant $t_\text{avg}$-range.

\begin{figure}
\includegraphics[width=0.4\textwidth]{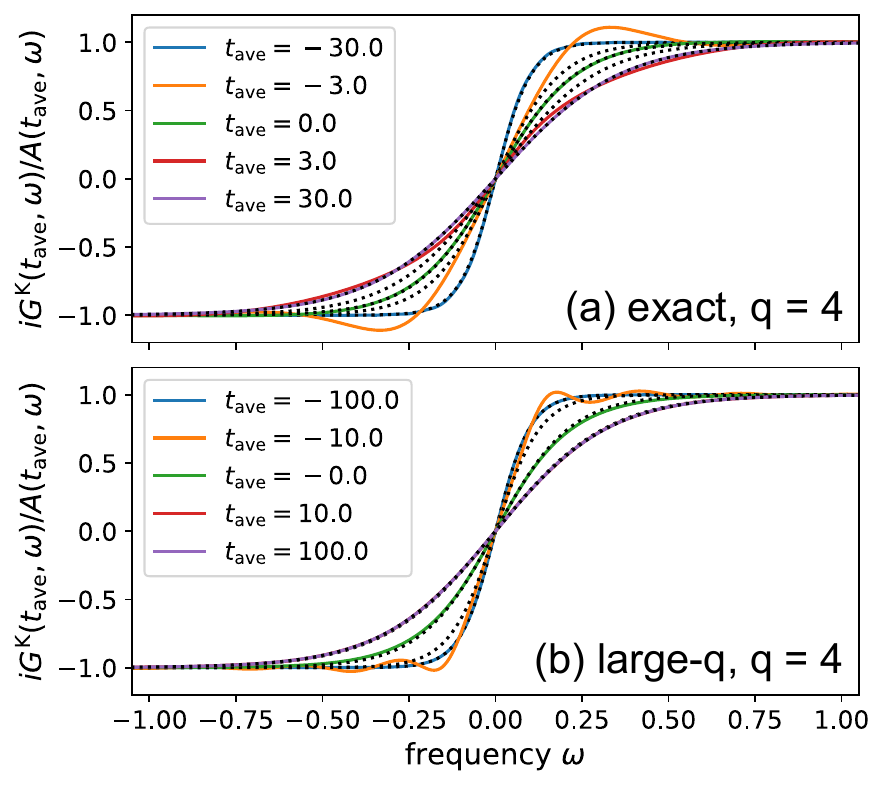}
\caption{Exemplary fits of the effective temperature from the ratio of Keldysh Green's function and spectral function\label{appfig:efftemps_fits}}
\end{figure}

In addition, the extraction of the effective temperature is susceptible to the choice of the frequency range over which the fit is performed.
At average times far away from the quench at $t_\text{avg} = 0$, we typically find that the ratio $i G^\text{K}(t_\text{avg}, \omega) / A(t_\text{avg}, \omega)$ coincides well with a thermal distribution function, justifying the concept of a thermalization rate of effective temperatures.
Close to the quench, the distribution functions are clearly nonthermal, as can be seen in Fig.~\ref{appfig:efftemps_fits} for both the full numerical solution and the large-$q$ theory.
We therefore only use a few frequency points around $\omega = 0$ for the fit to $i G^{\text{K}}(t_\text{avg}, \omega) / A(t_\text{avg}, \omega) \overset{!}{=} \tanh(\beta\omega/2) \simeq \beta\omega/2$.

\begin{figure}
\includegraphics[width=0.35\textwidth]{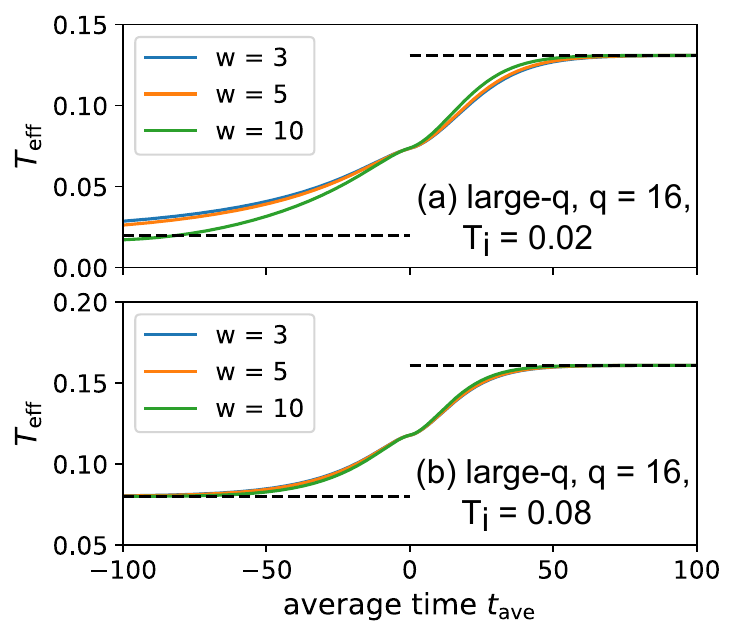}
\caption{Exemplary dependence of the numerically extracted effective temperature $T_\text{eff}$ on the choice of the fitting range: we fit the temperature from a number of $w$ frequency points with a spacing of $\Delta\omega = 2 \cdot 10^{-3}$ around $\omega = 0$. large-q, $q = 16$.\label{appfig:efftemps-fitrange}}
\end{figure}

As a quality indicator for all these choices, we make sure that the initial and expected final temperatures are correctly reproduced.
Fig.~\ref{appfig:efftemps-fitrange} demonstrates that, while final temperatures typically do not deviate much, the obtained initial temperatures at large values of $q$ may depend on numerical details like the width of the frequency window over which the temperatures are fitted.
In such cases, we make sure to work with data sets in which the numerically obtained initial temperature coincides with the expected one.

\begin{figure}
\includegraphics[width=0.35\textwidth]{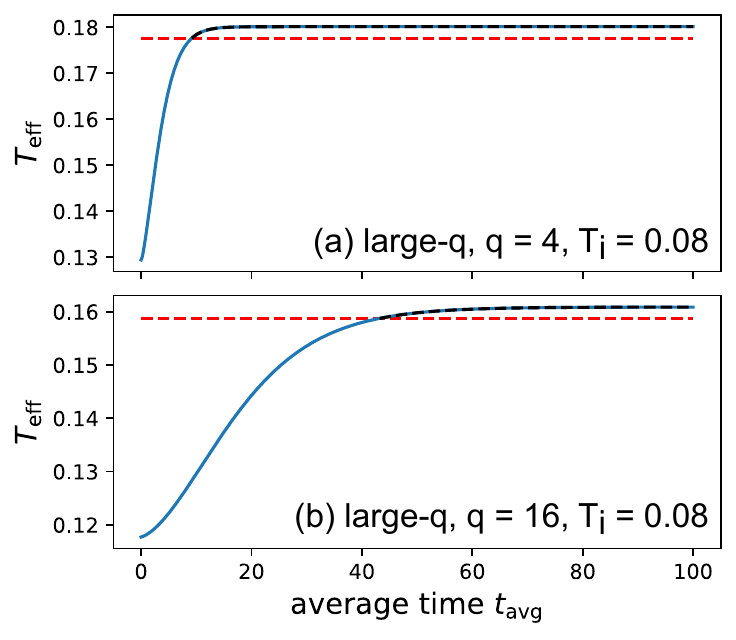}
\caption{Procedure to determine the thermalization rate $\Gamma$ from the effective temperature data. The dashed orange line indicates $95\%$ ($\epsilon = 0.05$) of the $T_\text{eff}$-range between the largest (final) value of $T_\text{eff}$ and the value at $t_\text{avg} = 0$.\label{appfig:thermrates-fit}}
\end{figure}

Lastly, thermalization rates $\Gamma$ need to be extracted from the effective temperature data.
In order to be able to compare rates obtained from different data sets, we establish a procedure how to extract the rates:
Taking the value span $\Delta T_\text{eff} = T_\text{eff}(t_\text{avg} \rightarrow \infty) - T_\text{eff}(t_\text{avg} = 0)$ of $T_\text{eff}$ from average time $t_\text{avg} = 0$ to the latest accessible average time in our simulations, we take a fraction $\epsilon$ of this range and fit an exponential decay
\begin{align}\begin{split}
 T_\text{eff}(t_\text{avg}) = T_f + \alpha \text{e}^{-\Gamma t_\text{avg}} .
\end{split}\end{align}
in the value range $\big[ (1-\epsilon) \Delta T_\text{eff}, \Delta T_\text{eff} \big]$.
This is demonstrated exemplarily in Fig.~\ref{appfig:thermrates-fit}.
We did not notice a strong dependence of the extracted rates on $\epsilon$ if $\epsilon \leq 0.05$.
For the plots in the main text we used $\epsilon = 0.01$.

\section{Full quench solution}
\label{fullSol}

In this appendix we restate the novel SYK solution derived in the companion paper~\cite{CompanionLouw2026}, namely the full Green's function solution for all four time blocks over the quench we consider
\begin{equation}
   \vec{\Jj}(t) = (2,2 + 2\gamma\Theta(t)), \quad \Jj(t) = \vert\vert \vec{\Jj}(t) \vert\vert  \label{quenchProtocol}
\end{equation}
where we keep $\Jj$ explicit, since it is needed to track the overall energy scale in App.~\ref{appSmallestTf}. For the quench we consider, and the system at half filling, the general solution is given by $\Gg^>(t_1,t_2) = -\imath e^{g(t_1,t_2)/q}/2$, where for the instantaneous thermal regions $t_1,t_2 >0$ ($(1,1)$ block) or $t_1,t_2 <0$ ($(0,0)$ block), we have the standard large $q$ solutions
	\begin{equation}
	e^{g_{b}(t)/2} = \frac{\cos(\pi v_{b}/2)}{\cos(\pi v_{b}/2- \i \lambda_{L,b} t/2)}, \label{expg}
	\end{equation}
with $s\in \{i,f\} \longleftrightarrow \{0,1\}$ and
\begin{equation}
\lambda_{L,b}  = 2\pi v_{b} T_{b}, \quad T_{b} = \frac{\Jj_{b} \cos(\pi v_{b}/2)}{\pi v_{b}}
\end{equation}
where we have defined $\vec{\Jj}_0 \equiv \vec{\Jj}(0^-)$ and $\vec{\Jj}_1 \equiv \vec{\Jj}(0^+)$ and their norms denoted without the arrow. The temperatures can be related by noting the relationship between post-/pre- quench Lyapunov exponents related via the expression
	\begin{equation}
		v_{f} = \frac{2}{\pi} \sin^{-1}\left[c_{01} \sin(\pi v_{i}/2)\right] \label{vrel}
	\end{equation}
where we have defined
\begin{equation}
    c_{01} \equiv \frac{\vec{\Jj}_0 \cdot \vec{\Jj}_1}{\Jj_{0} \Jj_{1}}
    = \frac{2+\gamma}{\sqrt{2\left(2+2\gamma+\gamma^{2}\right)}}.
    \label{CifDef}
\end{equation}

The off-diagonal block $t_1 < 0 ,t_2 > 0$, is inherently NEQ
\begin{align} g_{\text{neq}}(t_1,t_2) =& g_{1}^*(t_2)+g_{0}(t_1) \notag\\&-2\ln\left[1+c_{01} V_{0}(t_1)V_{1}(-t_2)\right]\label{gbnew}
\end{align}
where 
	 \begin{equation}
	 V_{b}(t) \equiv\frac{2 \Jj_{b}}{\lambda_{L,b}\coth(\lambda_{L,b} t/2) + 2 \i\Jj_{b} \sin(\pi v_{b}/2)  }.
	 \end{equation} 

The remaining block $t_2 < 0 < t_1$, the $(1,0)$ block probed in Sec.~\ref{secblocks}, follows from the conjugation relation $g(t_2,t_1)^{*} = g(t_1,t_2)$.

\subsection{Smallest final temperature}
\label{appSmallestTf}

Consider the protocol \eqref{quenchProtocol} so that the ratio controlling \eqref{vrel} is $c_{01}$ as defined in \eqref{CifDef}.
We wish to find the smallest final temperature  
\begin{equation}
    T_f = \frac{\Jj_1 \cos(\pi v_f/2)}{\pi v_f}
\end{equation}
attainable from a quench of strength $\gamma$, assuming a small initial temperature $T_i$.  For this the initial Lyapunov exponent is 
    \begin{equation}
        v_i = 1 - 2 T_i/\Jj_0 + \Oo(T_i^2).
    \end{equation}
Using \eqref{vrel}, the final temperature is given by
    \begin{align*}
    T_f \sim&  T_f(T_i=0)+ T_f(T_i=0)\\&\cdot \frac{\pi ^2 (\gamma +2) (T_i/\Jj)^2 \left(\gamma/2 +\sqrt{2} T_f(T_i=0)/\Jj+1\right)}{2 \gamma ^2}
    \end{align*}
    where 
    \begin{equation}
        T_f(T_i=0) = \frac{\Jj_1 \sqrt{1-c_{01}^2}}{2 \sin^{-1}\left[c_{01}\right]}. 
    \end{equation}

\section{Re-summation of the large-$q$ solution} \label{appResum}

In this appendix we discuss the effect of a mild rescaling of time in the analytic large-$q$ solution for the Green's function discussed in App.~\ref{fullSol}, and demonstrate that this modification does not change the order of the comparison between the analytic and numerical results.

When comparing large $q$ derived results with finite $q$, it is useful to consider some renormalized quantities that change variables at higher order in $1/q$. Like this, the result at leading order remains the same, but higher-order terms reflect the finite $q$ results better. For instance we make the ansatz that the Green's functions can be written as (at half filling)
\begin{equation}
    \Gg(t_1,t_2) = -\imath\,\text{sgn}(t_1-t_2)e^{g(t_1,t_2)/q}/2.    
\end{equation}

Note that in the above we have an expression formally to all order in $1/q$. This should be seen as a resummation of the finite $q$ Green's functions which works particularly well at low temperatures. The reason for this is that together with a renormalized coupling constant $\Jj$ 
\begin{equation}
    \Jj' = \frac{\Jj}{\sqrt{\left(1-\frac{2}{q}\right) \frac{\tan (\pi/q)}{\pi/q}}} \label{renorm_coupling}
\end{equation}
the solution would then exactly match the low temperature conformal result at all $q$ \cite{Maldacena2016Nov}. This solution is however only for late time and low temperature, as such, we will still find deviations at any finite $q$ at early times even if we formally took the zero temperature limit.

In our case we wish to show that the difference between the full numerical Green's function at finite $q$, and the analytical prediction using a $1/q$ expansion $\Gg_{\text{eq}}^{(\text{ana})}(t_1,t_2)$ are equal up to order $1/q^2$. In other words, we wish to show that the measure
\begin{equation}
d(q) = \max\limits_{\substack{t_1,t_2}}
    |\Gg(t_1,t_2)-\Gg_{\text{eq}}^{(\text{ana})}(t_1,t_2)| = \Oo(1/q^2) \label{dq}
\end{equation}
while recognizing that at moderate $q$ this equivalence is often obscured by finite-size effects in the numerics. Now we already know that all time dependence in $\Gg_{\text{eq}}^{(\text{ana})}$ is of order $1/q$. 

We find that these finite-size distortions are significantly reduced if, instead of directly comparing $\Gg$ and $\Gg^{(\mathrm{ana})}$, we introduce a rescaled analytic correlator
\begin{equation}
\Gg^{(\mathrm{ana})}_r(t_1,t_2)
\equiv \Gg^{(\mathrm{ana})}\!\left(\frac{t_1}{r}, \frac{t_2}{r}; q\right),
\qquad
r = 1 - \frac{6}{q^2}.
\label{eq:rescaled_analytic}
\end{equation}
To assess the formal impact of this rescaling, we expand $1/r = 1 + 6/q^2 + \Oo(1/q^4).$ Therefore, the rescaling-induced correction satisfies
\begin{equation}
\Gg^{(\mathrm{ana})}_r - \Gg^{(\mathrm{ana})}
= \Oo(t/q^3),
\end{equation}
where $t$ denotes a characteristic time scale in the correlator. For time arguments in the standard large-$q$ validity regime $1 \ll \Jj t \ll q$, these two are thus asymptotically equivalent.

Hence, the rescaling in Eq.~\eqref{eq:rescaled_analytic} does not alter the equivalence between $\Gg$ and $\Gg^{(\mathrm{ana})}$ up to order $1/q$; it only reorganizes higher-order $\Oo(1/q^3)$ contributions.

Practically, applying the rescaling of Eq.~\eqref{eq:rescaled_analytic} leads to a pronounced suppression of finite-size artifacts at moderate $q$. After this adjustment, the scaled deviation still satisfies \eqref{dq}, and also exhibits a clean exponential decay in $q$, revealing the expected asymptotic behavior and confirming that the analytic and numerical Green's functions coincide up to order $1/q$.

This rescaling can be viewed as an infinitesimal time reparametrization, consistent with the subleading corrections that appear in large-$q$ analyses of the SYK model. It effectively removes small $O(1/q^3)$ distortions without modifying the leading-order structure. 

\end{document}